\documentclass[preprint2, times, twocolappendix]{aastex631}

\newcommand{\projecttitle}{Gnuastro: Generating polar plots in Astronomical images}
\newcommand{\projectversion}{5d34243}
\newcommand{\projectgitbranch}{polar-plot}
\newcommand{\projectgitrepo}{https://codeberg.org/gnuastro/papers.git}
\newcommand{\projectcopyrightowner}{Mohammad Akhlaghi <mohammad@akhlaghi.org>}
\newcommand{\gnuastroversion}{0.22.69-ac1c}

\newcommand{\maneageversion}{8161194}

\newcommand{\downsample}{23.9}

\newcommand{\figradiussmall}{112}
\newcommand{\figradiuslarge}{150}
\newcommand{\figangletotal}{360}
\newcommand{\figanglesmall}{45}
\newcommand{\figanglelarge}{135}

\ifdefined\highlightnew

\else

\fi

\ifdefined\highlightnotes
\newcommand{\tonote}[1]{\textcolor{red!60!black}{[#1]}}
\else
\newcommand{\tonote}[1]{{}}
\fi

\usepackage{graphicx}

\usepackage{xcolor}
\color{black} 
\definecolor{DarkBlue}{RGB}{0,0,90}

\hypersetup{
    pdftitle={\projecttitle},
    pdfauthor={\projectcopyrightowner},
    pdfsubject={\projectgitrepo{} (commit \projectversion)},
    pdfkeywords={Reproducible research, Maneage, ADD YOUR OWN}
}

\usepackage{tikz}
\usetikzlibrary{external}
\tikzsetexternalprefix{tex/tikz/}

\usepackage{pgfplots}
\pgfplotsset{compat=newest}
\usepgfplotslibrary{groupplots}
\pgfplotsset{
  axis line style={thick},
  tick style={semithick},
  tick label style = {font=\footnotesize},
  every axis label = {font=\footnotesize},
  legend style = {font=\footnotesize},
  label style = {font=\footnotesize}
  }

\definecolor{smallangle}{rgb}{0.278,0.122,0.479}
\definecolor{largangle}{rgb}{0.133,0.461,0.566}
\definecolor{smallradius}{rgb}{0.199,0.822,0.222}
\definecolor{largradius}{rgb}{1,0.910,0.141}

\shorttitle{\projecttitle}
\shortauthors{Eskandarlou, S}

\begin{document}

\title{\projecttitle}

\author[0000-0002-6672-1199]{Sepideh Eskandarlou}
\affiliation{Centro de Estudios de F\'isica del Cosmos de Arag\'on (CEFCA), Plaza San Juan 1, 44001 Teruel, Spain}

\correspondingauthor{Sepideh Eskandarlou}
\email{sepideh.eskandarlou@gmail.com}

\author[0000-0003-1710-6613]{Mohammad Akhlaghi}
\affiliation{Centro de Estudios de F\'isica del Cosmos de Arag\'on (CEFCA), Plaza San Juan 1, 44001 Teruel, Spain}

\begin{abstract}
  \noindent
    When a structure displays dependence on distance and azimuthal angle from a center (for example the spiral arms of galaxies or the diffraction spikes of stars), projecting the pixels to polar coordinates greatly simplifies their study.
  This projection from one pixel grid to another is known as a ``polar plot''.
  For this purpose, a new option has been added to the GNU Astronomy Utilities (Gnuastro) in version 0.23 to ``\texttt{astscript-radial-profile}'' script, which we describe in this research note.
  The figures of this research note are reproducible with Maneage, on the Git commit \projectversion.
\end{abstract}


\section{Introduction}\label{sec:intro}
\noindent

Studying or identifying structures that exhibit polar symmetries can be significantly enhanced when projecting their pixels into polar coordinates (polar angle vs. radius).
For instance, to study the nature and mechanism of the age distribution across M101 and its spiral arm structure, \citet{garner24} used a polar plot.
Moreover, truncation in face-on galaxies is challenging to detect because these galaxies lose their circular symmetry and become asymmetrical towards their faint outer regions.
\citet{peters17} used the polar plot to study the disc truncation and extended halos in face-on spiral galaxies.
Furthermore, the Point Spread Function (PSF) of most telescopes exhibits spikes, and most surveys mask all the star spikes entirely \citep[for example, see Figures 7 and 11 of][]{coupon18}.
Polar plots offer a convenient way to identify their azimuthal angle (they rotate against the background sky).
Additionally, \citet{fang23} used polar plots (or ``polar-coordinate transformation'') to preprocess galaxy images in their neural network for rotationally-invariant galaxy morphology classification.

The new \texttt{--polar} option has been added recently to the \texttt{astscript-radial-profile} script \citep{infante24}, allowing straightforward generation of a polar plot on a given region of an image as described in the following section.

This feature has been included in GNU Astronomy Utilities, or Gnuastro\footnote{\url{https://www.gnu.org/software/gnuastro/manual/}} \citep{akhlaghi15}, since version 0.23\footnote{This paper was published before the release of Gnuastro v0.23. If it is not yet available, please use the latest test alpha release.}.

\newpage
\section{Generating the Polar Plot}\label{sec:analysis}
\begin{figure*}[!t]
  \begin{center}
  \ifdefined\makepdf%
    \tikzsetnextfilename{fig-polar-plot}%
    \input{tex/src/fig-polar-plot.tex}%
  \else
    \includegraphics[width=\linewidth]{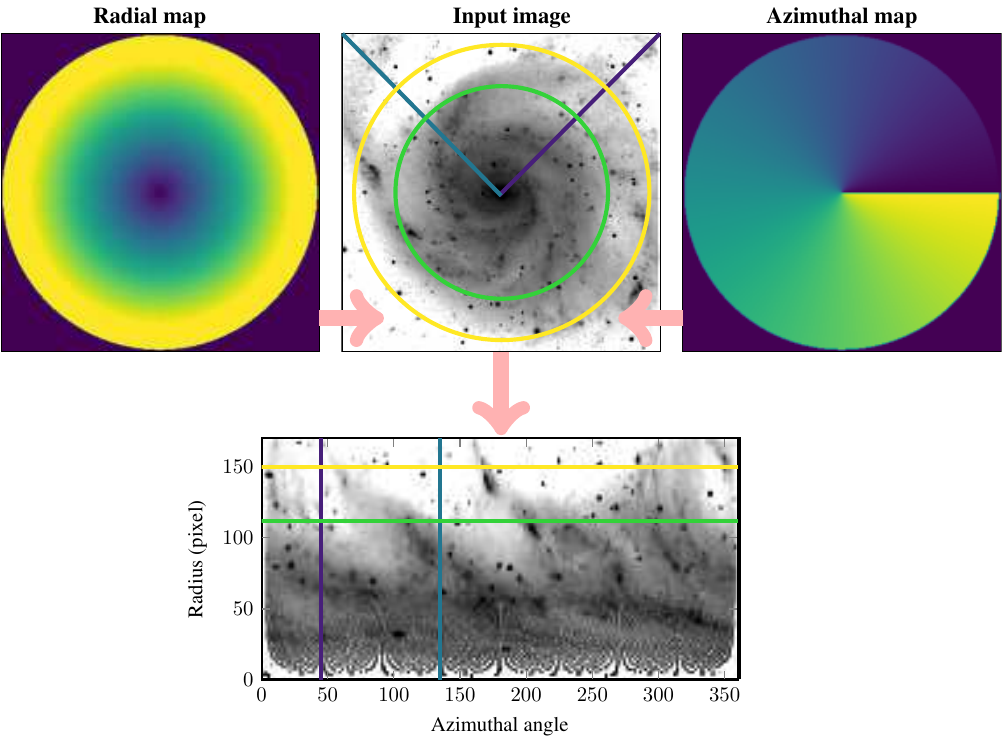}
  \fi

  \caption{\label{fig:image-histogram} Construction of polar plots.
    The top row, shows the radial map, input image (M101 as imaged in J-PLUS) and azimuthal map (from left to right).
    In radial and azimuthal maps, the color changes from dark blue to yellow as the angle and radius become larger.
    The combination of the azimuthal angle, input image, and radial profile generates the polar plot displayed in the bottom row.
    The horizontal represents the angles derived from the azimuthal data.
    As a demonstration, the dark purple and blue lines in the top-middle panel show the pixels at a constant azimuthal angle of {$\figanglesmall$} and {$\figanglelarge$} degrees respectively.
    The vertical axis represents the radii derived from the radial profile, and the value of each pixel corresponds to the value of the input image's pixels.
    As a demonstration, the green and yellow circles in the top-middle panel show the pixels at a constant radius of {$\figradiussmall$} and {$\figanglelarge$} respectively.}
  \end{center}
\end{figure*}

The \texttt{astscript-radial-profile} is an installed script that utilizes several lower-level Gnuastro programs to obtain the radial profile.
To create a polar plot, it performs the following steps:

\begin{itemize}
  \setlength\itemsep{-1mm}
\item Gnuastro’s MakeProfiles is used to create a radial map (where each pixel's value is its radius from the given center, top-left panel of Figure~\ref{fig:image-histogram}).
  This step is performed within the script, even if a polar plot is not requested.
  \item MakeProfiles is called a second time, but with the goal of generating the azimuthal map (where each pixel shows its azimuthal angle around the center on any elliptical shape, top-right panel of Figure~\ref{fig:image-histogram} shows it for a circle).
  \item To combine the two labeled images (representing the radius and azimuth of each pixel) with the pixel values of the target image, we use MakeCatalog.
    The feature that MakeCatalog in Gnuastro offers, which other catalog-based tools do not, is the ability to specify two labeled images as input when generating the catalog, see  \citet{akhlaghi19}.
    Finally, MakeProfiles is called one last time to create the polar plot based on the outputs of MakeCatalog.
    The resulting output is displayed in the bottom row of Figure~\ref{fig:image-histogram}.
\end{itemize}

A visual representation of all the steps of how polar plot construction are illustrated in Figure~\ref{fig:image-histogram}.
The top-middle panel of Figure~\ref{fig:image-histogram} shows M101 as imaged in J-PLUS DR3 in the rSDSS filter \citep{cenarro19}; down-sampled to {$\downsample$}$^{\prime\prime}$/pix.
In the radial profile image, with dark blue colors indicate smaller radii.
As the profile extends outward and the color transitions to yellow, the radii increase.
This color progression effectively illustrates the change in radius from the center to the outer edges of each pixel of the input image in the vertical axis of polar plot.
The green circle/line in the input/polar images represent a radius of {$\figradiussmall$}, while the yellow circle/line represents a radius of {$\figradiuslarge$}.
These radii are extracted from the radial profile, providing clear markers for the specified radii of input image.

Similarly, in the azimuth panel, the dark blue color represents a smaller angle (zero degrees).
The angle then increases by one degree in a counterclockwise direction, culminating in yellow, which indicates a {$\figangletotal$}-degree rotation.
This color gradient depicts the full range of azimuthal angles within the panel for each pixel of input image in x-axis of polar plot.
In the input image and polar plot, the purple line indicates an angle of {$\figanglesmall$} degree, while the blue line represents an angle of {$\figanglelarge$} degree.
The angles are extracted from the azimuth angle image, providing a clear visual reference for their respective angle in input image and polar plot.
The final output, which is the polar plot resulting from combining the three panels in the top row, is displayed in the bottom row of Figure~\ref{fig:image-histogram}.

\section{Acknowledgement}

The workflow of this research note uses Maneage \citep[\emph{Man}aging data lin\emph{eage},][commit \maneageversion]{maneage}.
This research note is created from the Git commit {\projectversion}, hosted on Codeberg\footnote{Git repository of the paper (\texttt{\small{\projectgitbranch}} branch): \url{\projectgitrepo}.} which is archived on Software Heritage\footnote{Software Heritage (SoftWare Hash IDentifier, SWHID): \href{https://archive.softwareheritage.org/swh:1:dir:4e09bf85f9f87336fa55920bf67e7bcf6d58bbd5;origin=https://codeberg.org/gnuastro/papers;visit=swh:1:snp:557ee1a90de465659659ecc46df0c5ce29d0bb61;anchor=swh:1:rev:375e12e52080006be6a28e10980e79ef54d13d1d}{swh:1:dir:4e09bf85f9f87336fa55920bf67e7bcf6d58bbd5}} for longevity.
Supplements are also available on Zenodo\footnote{Zenodo: \href{http://doi.org/10.5281/zenodo.11403643}{11403643}}.
The analysis uses GNU Astronomy Utilities \citep[Gnuastro,][]{akhlaghi15,akhlaghi19} v\gnuastroversion.
Gnuastro has been funded by the Japanese MEXT scholarship and its Grant-in-Aid for Scientific Research (21244012, 24253003), ERC 339659-MUSICOS, Spanish MINECO AYA2016-76219-P, and NextGenerationEU ICTS-MRR-2021-03-CEFCA.
We acknowledge the funding by Governments of Spain and Arag\'on through FITE and Science Ministry (PGC2018-097585-B-C21 and PID2021-124918NA-C43).

\newpage
\bibliography{references}{}
\bibliographystyle{aasjournal}

\appendix

\section{Software acknowledgement}
\label{appendix:software}
 
This research was done with the following free software programs and libraries:  1.23, Bzip2 1.0.8, CFITSIO 4.1.0, CMake 3.24.0, cURL 7.84.0, Dash 0.5.11-057cd65, Discoteq flock 0.4.0, Expat 2.4.1, File 5.42, Fontconfig 2.14.0, FreeType 2.11.0, Git 2.37.1, GNU Astronomy Utilities 0.22.69-ac1c \citep{gnuastro,akhlaghi19}, GNU Autoconf 2.71, GNU Automake 1.16.5, GNU AWK 5.1.1, GNU Bash 5.2-rc2, GNU Binutils 2.39, GNU Bison 3.8.2, GNU Compiler Collection (GCC) 12.1.0, GNU Coreutils 9.1, GNU Diffutils 3.8, GNU Findutils 4.9.0, GNU gettext 0.21, GNU gperf 3.1, GNU Grep 3.7, GNU Gzip 1.12, GNU Integer Set Library 0.24, GNU libiconv 1.17, GNU Libtool 2.4.7, GNU libunistring 1.0, GNU M4 1.4.19, GNU Make 4.3, GNU Multiple Precision Arithmetic Library 6.2.1, GNU Multiple Precision Complex library, GNU Multiple Precision Floating-Point Reliably 4.1.0, GNU Nano 6.4, GNU NCURSES 6.3, GNU Readline 8.2-rc2, GNU Scientific Library 2.7, GNU Sed 4.8, GNU Tar 1.34, GNU Texinfo 6.8, GNU Wget 1.21.2, GNU Which 2.21, GPL Ghostscript 9.56.1, Help2man , Less 590, Libffi 3.4.2, Libgit2 1.3.0, libICE 1.0.10, Libidn 1.38, Libjpeg 9e, Libpaper 1.1.28, Libpng 1.6.37, libpthread-stubs (Xorg) 0.4, libSM 1.2.3, Libtiff 4.4.0, libXau (Xorg) 1.0.9, libxcb (Xorg) 1.15, libXdmcp (Xorg) 1.1.3, libXext 1.3.4, Libxml2 2.9.12, libXt 1.2.1, Lzip 1.23, OpenSSL 3.0.5, PatchELF 0.13, Perl 5.36.0, pkg-config 0.29.2, podlators 4.14, Python 3.10.6, util-Linux 2.38.1, util-macros (Xorg) 1.19.3, WCSLIB 7.11, X11 library 1.8, XCB-proto (Xorg) 1.15, xorgproto 2022.1, xtrans (Xorg) 1.4.0, XZ Utils 5.2.5 and Zlib 1.2.11. 
The \LaTeX{} source of the paper was compiled to make the PDF using the following packages: courier 61719 (revision), epsf 2.7.4, etoolbox 2.5k, helvetic 61719 (revision), lineno 5.3, pgf 3.1.10, pgfplots 1.18.1, revtex4-1 4.1s, tex 3.141592653, textcase 1.04 and ulem 53365 (revision). 
We are very grateful to all their creators for freely  providing this necessary infrastructure. This research  (and many other projects) would not be possible without  them.

\end{document}